\documentclass[epj]{svjour}

\usepackage{graphicx}

\journalname{European Physics Journal}

\begin{document}

\title{Effects of the Screening Breakdown in the Diffusion-Limited Aggregation Model}

\author{S. C. Ferreira Jr.\\ \email{silviojr@ufv.br}}

\institute{Departamento de F\'{\i}sica, Universidade Federal de Vi\c{c}osa, 36571-000, Vi\c{c}osa,
Minas Gerais, Brazil}

\abstract{Several models based on the diffusion-limited aggregation (DLA) model were proposed and their scaling properties explored by computational and theoretical approaches. In this paper, we consider a new extension of the on-lattice DLA model in which the unitary random steps are replaced by random flights of fixed length. This procedure reduces the screening for particle penetration present in the original DLA model and, consequently, generates new pattern classes. The patterns have DLA-like scaling properties at small length of the random flights.  However, as the flight size increases, the patterns are initially round and compact but become fractal for sufficiently large clusters. Their radius of gyration and number of particles at the cluster surface scale asymptotically as in the original DLA model.  The transition between compact and fractal patterns is characterized by wavelength selection, and $1/k$ noise was observed far from the transition.}

\date{\today}

\titlerunning{\sc Effects of Screening\ldots}
\authorrunning{\sc S. C. Ferreira}
\maketitle

\section{Introduction}

\label{intro}

Pattern formation is a field of great interest in the Non-equilibrium Statistical Physics. In special, the diffusion-limited aggregation (DLA) model \cite{Witten} is a noteworthy example in which a very simple algorithm generates complex disorderly patterns. This model was related to several physical and biological applications, such as electrodeposition \cite{Matsushita}, viscous fingering \cite{Maloy}, bacterial colonies \cite{Matsushita2}, neurite formation \cite{Caserta}, and, more recently, tumor growth \cite{Sander,jr}. After two decades, the DLA scaling properties yet remain not completely understood and this model constitutes an actual and challenger theoretical problem. In the original model \cite{Witten}, the simulation begins with an initial seed at the center of the lattice. A particle, represented by a site of the lattice, is released at a random distant point (the launching radius) and performs an on-lattice random walk. If the particle visits a neighbor site of the initial seed, it joins irreversibly to this site. If the distance of the particle to the cluster is too large, the particle is excluded and a new one is released from the launching circle. Successive new particles are released from a new random distant point and walk at random until they find the cluster neighborhood. This algorithm is inefficient and clusters containing just a few thousand particles can be generated with a considerable computational effort. However, much more efficient algorithms can be used in order to grow clusters with more than $10^8$ particles for lattices \cite{Ball,Tolman} and off-lattices \cite{Tolman,Kaufman} DLA models.

Due its importance as a fundamental model, several variants of the DLA model
were proposed \cite{Meakin0}. In special, models focusing its screening properties had remarkable interest. For example, Meakin investigated a model in which the random walks (trajectories with fractal dimension 2.0) of the DLA
model are replaced by fractal trajectories (Levy flights or Levy walks) \cite{Meakin1}. As the fractal dimensions of the trajectories decrease, the particles become more penetrating and, consequently, the DLA screening effects
less intensive. As a consequence, the fractal dimensions of the clusters increase and their branches become denser. Also, drift-diffusion-limited aggregation models, in which a particle follows radially biased random walks
toward \cite{Meakin2} and away \cite{Nagatani} from the initial seed particle, were considered. The former case generates DLA-like clusters on shorter length scales and circular-dense patterns, with fractal dimension 2.0 on longer length scales as the drift probability increases. The later case also generates DLA-like clusters on shorter length scales, but eccentric patterns are generated on longer length scales, with fractal dimensions close to 1.0. Moreover, in
both cases the patterns are sparse containing several internal holes. In order to generate viscous fingering patterns, several models  introducing surface tension and surface relaxation effects in the DLA model were proposed \cite{Vicsek,Meakin3,Tao,Fernandez}. In these models, a sticking probability dependent of the local curvature for the particles arriving at the surface of the aggregate is assumed. If the particle does not stick to the cluster, it continues diffusing. Again, the DLA screening is weakened since the particles do not stick with probability 1. Depending on the surface tension, these models generate Eden-like round and compact patterns \cite{Eden} on shorter length scales that become branched for asymptotically large clusters.

In this paper, we present a new extension of the lattice DLA model in which the unitary random steps are replaced by random flights of fixed length. This procedure allows the particles to penetrate the screening present in DLA clusters and, consequently, generates new pattern classes. In Sec. \ref{model}, the model and the computational algorithms are described. In Sec. \ref{results} the model results are presented and discussed. Finally, some conclusions are drawn in Sec. \ref{conclu}.

\section{Model and methods}
\label{model}

In the modified DLA model presented in this paper, the particles perform off-lattice random flights whose step length is given by $\Delta=\delta a$, where $a$ is the lattice constant (assumed unitary) and $\delta\ge 1$ is the unique model parameter. Notice that flights substitute walks and, consequently, at each step the particle occupies a random position in a hypersphere of radius $\delta$ centered on its initial position. Thus, depending on the $\delta$ value, a particle near to the border of the aggregate can reach its inner regions forbidden to particles executing normal random walks. For sake of simplicity, the particles of the aggregate lie on a regular square lattice. Two conditions determine when the flighting particles finish their trajectories becoming part of the aggregate. If a particle visits a nearest neighbor site of the aggregate, it sticks irreversibly to this site (like in the DLA model). But, if the particle reach a site belonging to the cluster, it realizes on-lattice random walks of unitary step until find an empty site, irreversible sticking on it. Here, a surface relaxation, also present in the sticking probability dependent models \cite{Vicsek,Meakin3,Tao,Fernandez}, was introduced. In order to determine when a particle visits a site, its lattice coordinates are defined as the nearest integers associated to its real coordinates. The particular case $\delta=1$ must reproduce the semi-lattice DLA model \cite{Meakin4}.

\begin{figure}[hbt]
\begin{center}
\resizebox{7cm}{!}{\includegraphics{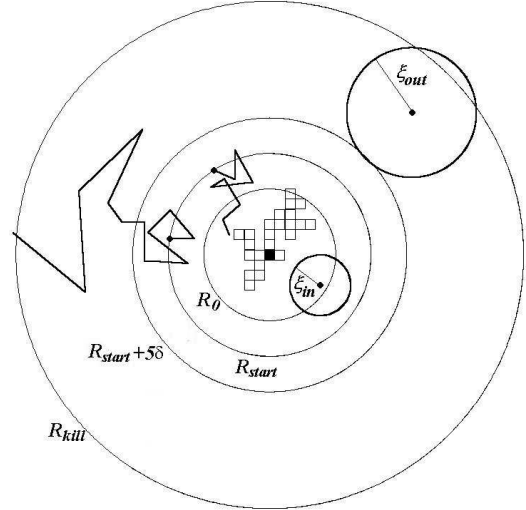}}
\end{center}
\caption{Optimized DLA model. All the radius defined in the algorithm are shown: the delimiting aggregate ($R_0$), the launching  ($R_{start}$) and sinking ($R_{kill}$) circles, as well as the circles that illustrate the long flights used when the particles are distant from the aggregate whether inside ($\xi_{in}$) or outside ($\xi_{out}$) the launching circle. The center of the lattice is depicted in black and the other sites of the aggregate are represented by white squares. Also, two trajectories, one toward the aggregate and the other in which the particle reaches the killing radius, are drawn.}
\label{dlamodel}
\end{figure}

To optimize the algorithm efficiency some procedures were adopted. The random trajectories started at a launching circle of radius $R_{start}=R_0+5\delta$ centered on the initial seed. Here, $R_0$ is the maximum distance of a site of the aggregate from the center of the lattice and the quantity $5\delta$ warranties that the particle realizes some steps before it reaches the aggregate. This procedure is justified since a random walker released far outside the launching circle will first intercept it at a random position with equal probability \cite{Meakin0}. A killing radius $R_{kill}=100\delta R_0$, proportional to the flight length, was defined. If a random walker cross the circle of radius $R_{kill}$ centered on the initial seed it is discarded. Also, we used a standard method to speed up the simulations \cite{Meakin0}, which consists in allow the particles outside the launching circle take long steps $\xi_{out}$ (longer than $\delta$). However, these long steps cannot bring up the particle to inside the launching circle. Thus, the step length was chosen as $\xi_{out}= \max (R-R_{start}-5\delta,\delta)$, where $R$ is the distance of the random walker from the center of the lattice. Following the last definition, one can see that $\xi_{out}>\delta$ only for $R>R_{start}+5\delta$.  In addition, even inside the cluster delimiting circle exit large empty regions. So, we adapted a algorithm used by Ball and Brady \cite{Ball}, in which the random walker jump a distance $\xi_{in}$ if the region delimited by the circle of radius $\xi_{in}$ centered on the walker does not contain occupied sites nor sites neighboring the cluster. Thus, the radius $\xi_{in}$ must define a region containing only sites from which the random walker cannot reach the aggregate by a single step. Therefore, $\xi_{in}$ was determined by the following procedure. The border of a sub-lattice of linear size $l=\max(40,R_0/10)$ centered on the random walker is visited. If none part of the aggregate was found, then the sub-lattice is empty and $\xi_{in}=l/2$ is adopted. Otherwise, if any part of aggregate was found, $l$ is reduced by a half and the whole procedure is repeated until found an empty sub-lattice or its size reaches the value $l<5\delta$. In former case a $\xi=l/2$ is assumed and in the latter a $\xi_{in}=\delta$ is adopted. Fig. \ref{dlamodel} shows a scheme of the methods described in this paragraph.

As the step length $\delta$ increases the patterns become denser and, consequently, the random walks over the aggregate are very computational time consuming. Thus, for $\delta > 20$, the random walkers over the occupied regions were also allowed to realize long steps if they are far away from the border.

\section{Results and Discussions}
\label{results}

\begin{figure}[hbt]
\begin{center}
\resizebox{8.0cm}{!}{\includegraphics{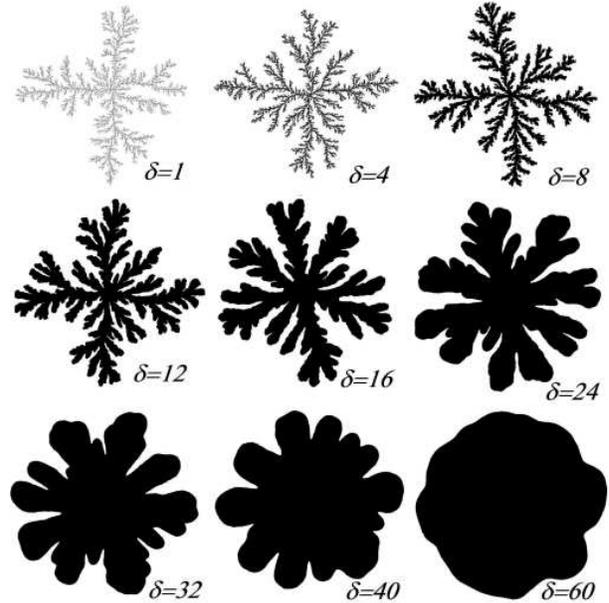}}
\end{center}
\caption{Growth patterns generated by the DLA model with screening breakdown. These clusters were generated in square lattices containing $2000 \times 2000$ sites. The aggregate with fewer particles ($\delta=1$) contains $1.5\times 10^5$ particles whereas the one with the larger number ($\delta=60$) contains $2.8 \times 10^6$ particles.}
\label{dlascr1}
\end{figure}

The DLA model with screening breakdown was simulated in square lattices containing up to $12000 \times 12000$ sites and  flight lengths ranging in the interval $1 \le \delta \le 100$. The maximum number of particles in the aggregates varied from $N\sim 10^6$, for smaller $\delta$ values, to $N\sim 10^8$, for the larger ones. In Fig. \ref{dlascr1}, growth patterns generated by the model with distinct $\delta$ values are shown. As the flight length increases from $\delta=1$, the branches of the clusters become denser and for larger $\delta$ values ($\delta=40$) compact patterns, with a nearly regular shape, emerge. In order to quantify the geometry of the clusters, their radius of gyration $R_g$ and the number of particles on the cluster border $S$ were evaluated. These quantities scale with the number of particles as
\begin{equation}
\label{rgscale}
R_g\sim N^\nu
\end{equation}
and
\begin{equation}
\label{sscale}
S \sim N^\sigma.
\end{equation}
Also, the fractal dimension is given by $d_f=1/\nu$.  As expected, the particular case $\delta=1$ reproduces the semi-lattice DLA model. The fractal dimension obtained from the slope $d\log N/d\log R_g$ is $d_f=1.69\pm 0.03$, a value smaller than $d_f=1.715\pm 0.004$ found for large off-lattice DLA simulations \cite{Tolman}. However, it is a well-understood fact that anisotropy effects of the square lattice reduce the effective fractal dimension of larger clusters \cite{Meakin4}. Indeed, simulations using a recent method proposed by Bogoyavlenskiy \cite{Bogoyavlenskiy} to avoid the lattice anisotropy for DLA clusters provide a fractal dimension $d_f=1.716 \pm 0.022$ for $\delta=1$ corroborating our previous claim.

New patterns emerge due to the weakness of the DLA screening resultant from the increasing of the flight lengths. This phenomenon can be qualitatively understood through an analogy with the sticking probability DLA models. In such models there is a correlation length $\lambda$ that increases as the sticking probabilities $P_{\sigma}$ decreases. Thus, the patterns are compact on smaller length scales ($\ell\ll\lambda$) but have DLA-like scaling properties at longer length scales ($\ell\gg\lambda$) \cite{Meakin0}. In order to use this argument for the DLA model with screening breakdown, we consider that the effective penetration depth of the particles in the sticking probability DLA models is proportional to $\sqrt{1/P_\sigma}$. Indeed, $1/P_\sigma$ is the mean number of steps realized by the particles in the nearness of the cluster border before they attach to the cluster and the mean square displacement of the random walker is proportional to the number of steps. This depth can be associated with the size of the flights $\delta$ and, consequently, the previous arguments deduced for the sticking probability problem can readily be extended to the present model.

\begin{figure}[hbt]
\begin{center}
\resizebox{8.5cm}{!}{\includegraphics{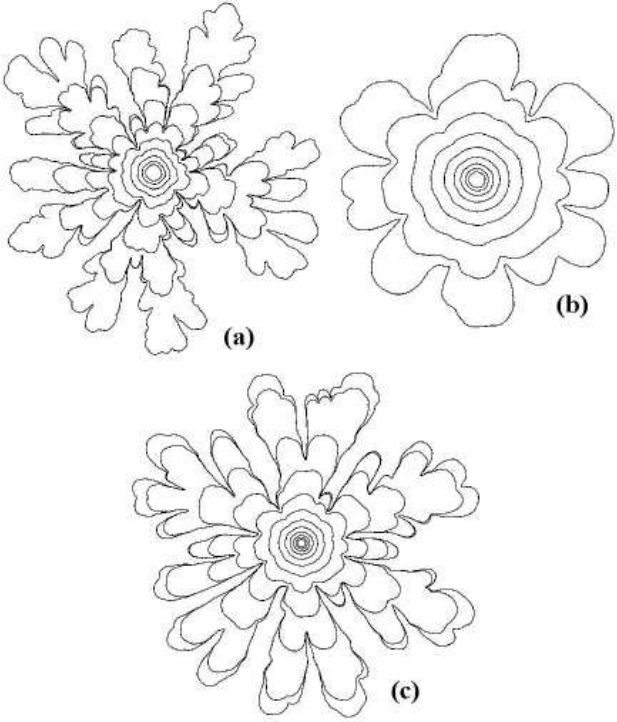}}
\end{center}
\caption{Cluster boundaries along the growth of the aggregate. The central boundary represents a cluster containing $10^4$ particles and the subsequent ones correspond to patterns after each cluster size doubling. Patterns with $5\times 10^6$ particles for (a) $\delta=20$ and (b) $\delta=40$ are shown. In (c) the continuation of the simulation for $\delta=40$ up to $3\times 10^7$ particles is shown. The figure (c) was reduced 50\% if compared with the figures (a) and (b).}
\label{dlascr2}
\end{figure}

The above mentioned length scales are neatly observed in Fig. \ref{dlascr2}, in which the surface sites of the growing clusters for distinct numbers of particles are drawn. As one can see, the more inner boundaries of Figs. \ref{dlascr2}(a) and (b) are initially Eden-like, exhibiting a round and rough border. As new particles are added to the clusters, their interfaces become progressively regular and, at a given stage, quasi-periodic finger-like structures are build up. At this stage, the patterns are featured by a characteristic wavelength spontaneously selected. However, this is a transient behavior because the ``sprouts'' consecutively split generating fractal structures at long times as illustrated in Fig. \ref{dlascr2}(c). In this figure, the sequence of the simulation shown in Fig. \ref{dlascr2}(b) after three doublings is drawn. Thus, any compact pattern becomes fractal for sufficiently large clusters. This behavior is analogue to that found in the DLA models with sticking probabilities \cite{Vicsek,Meakin3}, but the layer mechanisms are different from those involved in  such models. There, the local curvature of the surface, that is explicitly included in those models, seems to develop a essential rule, while in the present model only the screening breakdown is taken in to account.

\begin{figure}[hbt]
\begin{center}
\resizebox{8.8cm}{!}{\includegraphics{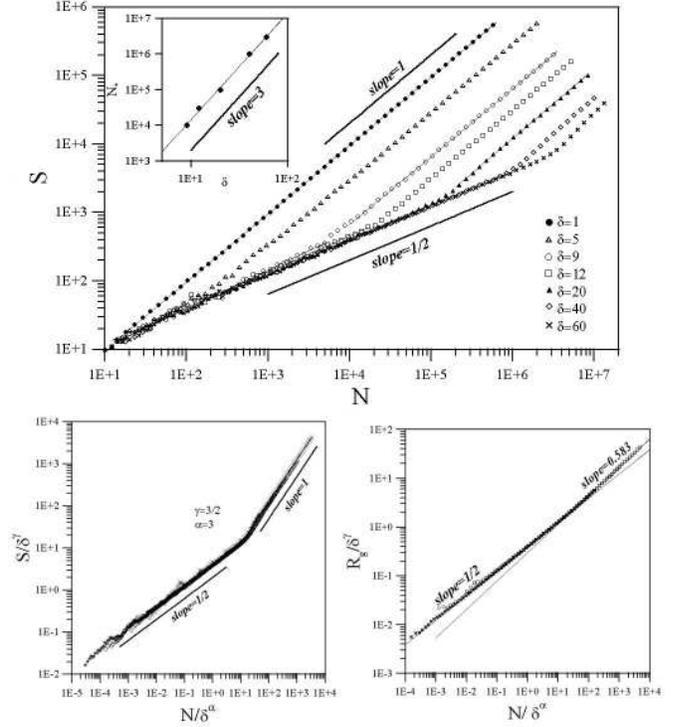}}
\end{center}
\caption{Radius of gyration $R_g$ and the number of peripheral sites $S$ as a function of the total number of particles $N$ in the cluster. In (a) the curves $S\times N$ are shown for different $\delta$ values. In (b) and (c) the collapsed curves of $S$ and $R_g$, respectively, are shown for $\delta$ ranging from $\delta=9.0$ up to $\delta=100$.}
\label{dlascr3}
\end{figure}

The radius of gyration $R_g$ and the number of peripheral particles $S$ are able to detect the crossover between Eden-like and DLA-like patterns suggested by Fig. \ref{dlascr2}. In Fig. \ref{dlascr3}(a), plots of $S$ as a function of the number of particles $N$ for distinct flight length sizes $\delta$ are shown. For small $\delta$ values an unique scaling law, in which $S\sim N$, is observed indicating that these patterns scale as the DLA model. As $\delta$ is increased, the curves exhibit two distinct slopes separated by a neat crossover. Below a characteristic number of particles $N_\times$, the clusters scale as the Eden model, i. e., $S\sim N^{1/2}$,  and scale as the DLA model above this value. The simulations show that for $\delta > 8.0$ $N_\times$ increases with $\delta$ as $N_\times \sim \delta^\alpha$, with $\alpha=3$. In the inset of Fig. \ref{dlascr3}(a), the curve $N_\times$ against $\delta$ and the correspondent power law fitting are shown. It is evident the quality of the fitting. For $\delta < 8.0$ the determination of the crossover becomes difficult. An identical crossover was observed for the radius of gyration. Moreover, the number of peripheral sites and the radius of gyration do not depend on the $\delta$ values for $N<N_\times$, as indicated in Fig. \ref{dlascr3}(a). However, above the crossover, both quantities increase as $\delta$ decreases. Indeed, we have that $S\sim \delta^{-z_1}$ and $R_g\sim \delta^{-z_2}$.  In figures \ref{dlascr3}(b) and \ref{dlascr3}(c) the curves $S/\delta^\gamma$ and $R_g/\delta^\gamma$ against $N/\delta^\alpha$, respectively, are plotted. The $\gamma$ exponent was varied in search for a very satisfactory collapse, and the best value is $\gamma\approx 3/2$. Using that the collapsed curves scale as in the Eden model for $N\ll N_{\times}$ and as in the DLA model for $N\gg N_{\times}$, the following scaling relationships are defined
\begin{eqnarray}
    S\sim N^{\frac{1}{2}}f\left(\frac{N^{\sigma-\frac{1}{2}}}{\delta^{\alpha\sigma-\gamma}}\right)\\
    R_g\sim N^{\frac{1}{2}}f\left(\frac{N^{\nu-\frac{1}{2}}}{\delta^{\alpha\nu-\gamma}}\right)
\end{eqnarray}
where $\sigma$ and $\nu$ are the exponents defined in Eqs. (\ref{rgscale}) and (\ref{sscale}) for the original DLA model, i. e., $\sigma=1$ and $\nu\approx0.583$. The scaling function $f(x)$ is defined as
\begin{equation}
f(x)=\left \{
    \begin{array}{l}
        const. \mbox{~if~} x \ll 1 \\
        x \mbox{~if~} x\gg 1
    \end{array}. \right.
\end{equation}
Thus, we have that $z_1=\alpha\sigma-\gamma=3/2$ and $z_2=\alpha\nu-\gamma\approx 0.249$.

\begin{figure*}[hbt]
\begin{center}
\resizebox{12cm}{!}{\includegraphics{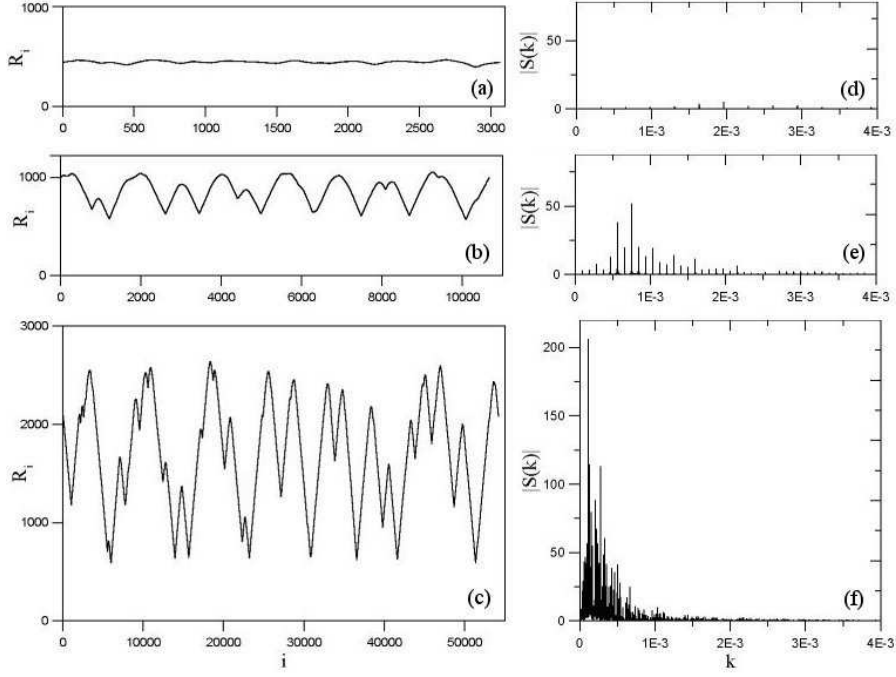}}
\end{center}
\caption{Profiles of (a) the last but two and (b) the last contours shown in Fig. \ref{dlascr2}(b). (c) Profile of the last contour of Fig. \ref{dlascr2}(c). The correspondent Fourier spectra are shown at right ((d)-(f)). The profiles are drawn in the same vertical scale whereas the horizontal ones depend on the contour. In turn, the Fourier spectra were plotted by using the same horizontal and vertical scales.}
\label{dlascr4}
\end{figure*}

One remarkable feature of the present model is the mode selection exhibited by the patterns at the compact to fractal transition illustrated in Figs. \ref{dlascr2} and \ref{dlascr3}. In order to analyze this feature, firstly we map the cluster contour in a $1+1$ profile, in which the heights represent the distance of a point in the contour from the center-of-mass of the pattern. Secondly, we calculated the Fourier spectra $S(k)$ of the profile by using the standard FFT method. Here, $k$ is the wavenumber and $1/k$ the correspondent wavelength. In Fig. \ref{dlascr4}, the mapped profiles of the contours drawn in Fig. \ref{dlascr2} and the correspondent Fourier spectra are shown. For the profile correspondent to the pattern before the crossover (Figs. \ref{dlascr4}(a) and \ref{dlascr4}(d)) no mode is selected and the observed low amplitude wavenumbers indicate the onset of the instabilities that will lead to the fingering shown in Fig. \ref{dlascr4}(b). The Fourier spectrum of the cluster border when $N\approx N_{\times}$ is characterized by the selection of a principal mode and additional lower-amplitude modes as indicated in Fig. \ref{dlascr4}(e). As the cluster increases, the selected wavenumber shifts to the left, indicating a larger wavelength selection, and additional modes emerge in the Fourier spectrum. This results are in qualitative agreement with fluid-fluid displacement experiments for radial fingering patterns \cite{rauseo} and recent experiments of grain-grain displacement in a Hele-Shaw cell \cite{couto}.

\begin{figure}[hbt]
\begin{center}
\resizebox{8cm}{!}{\includegraphics{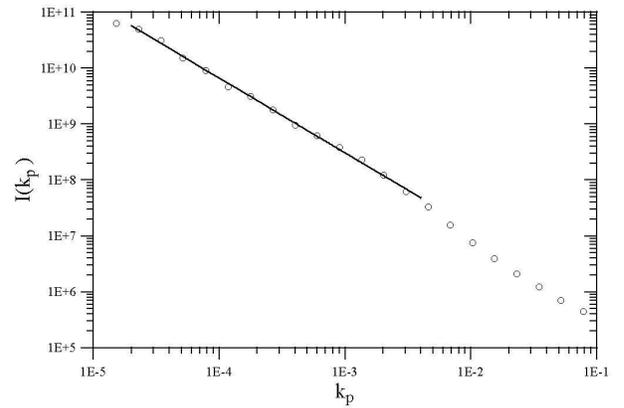}}
\end{center}
\caption{Typical integrated power spectrum for profiles generated from the pattern contours. The flight length used was $\delta=10$ range of and the fit is excludes very long and very short wavelengths.}
\label{dlascr5}
\end{figure}

In the DLA-like scaling regime ($N>>N_\times$) a wide range of wavelengths are selected as suggested in  Fig. \ref{dlascr4}(f). This indicates the presence of self-affine profiles characterized by $1/k^\zeta$ noise, i. e., all wavelengths are present in the Fourier spectra. Thus, the power spectrum of the profile scales as
\begin{equation}
|S(k)|^2\sim k^{-\zeta},
\end{equation}
where $\zeta=2H+1$ establishes the relationship between $\zeta$ and the Hurst exponent $H$ \cite{Meakin0}. Since the power spectra are noisy and the wavenumbers are, due the FFT regression, evenly spaced, it is hard to determine the power law exponent. Therefore, we were led to consider the band-integrated regression \cite{pilgram}, in which the power spectrum estimated by FFT is integrated over the wavenumber intervals $[k_0,\beta k_0]$, $[\beta k_0,\beta^2k_0]$, $[\beta^2k_0,\beta^3k_0]$, and so on. So, we computed the quantity
\begin{equation}
I(p)=\sum_{k=\beta^p k_0}^{\beta^{p+1}k_0} |S(k)|^2 \sim k_p^\eta,
\end{equation}
where $k_p=\beta^p k_0$ and $\eta\equiv 1-\zeta \equiv H/2$. In Fig. \ref{dlascr5}, the integrated power spectrum for  simulations with $\delta=10$ is shown. The $\eta$ exponent varies during the growth of the aggregate from the value $\eta=2$ ($H=1$), that characterizes a flat profile, to the value $\eta \approx 1.35$ ($H \approx 0.68$), indicating a positively persistent profile, far from the crossover. Our simulations were not able to determine the asymptotic $\eta$ values for all distinct $\delta$ values studied due to computational limitations (the maximum lattice size simulated was $12000 \times 12000$). However, simulations for $\delta \in [8,18]$ provide approximately the same asymptotic $\eta$ value and, consequently, we conjecture that this exponent is independent of $\delta$. It is important to notice that for $\delta < 8$ the pattern contours cannot be mapped the in  self-affine profiles because very thin branches are present and the sequential mapping algorithm used becomes undefined.

\begin{figure}[hbt]
\begin{center}
\resizebox{8cm}{!}{\includegraphics{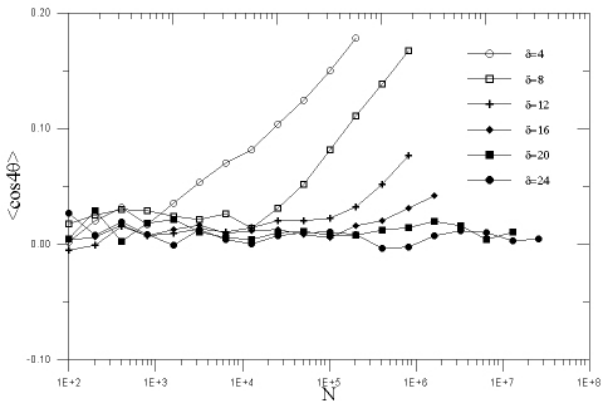}}
\end{center}
\caption{Mean values of the fourth circular harmonic for distinct $\delta$ values. The number of samples used on the averages varies from $100$ for $\delta=4$ to $30$ for $\delta=24$.}
\label{dlascr6}
\end{figure}

An important feature of the DLA model with screening breakdown is the formation of isotropic patterns for large $\delta$ values. This is in marked contrast with the anisotropic patterns generated for smaller $\delta$ values, a central feature of on-lattice DLA models \cite{Meakin0} (see Figs. \ref{dlascr1} and \ref{dlascr2}). In order to quantify the cluster anisotropy, we calculated the fourth circular harmonic $\langle \cos 4\theta\rangle$, where $\theta$ is the angular position of the particle measured from the initial seed position. The averages were done over all peripheral particles of up to $100$ samples. This procedure excludes the effects of internal compact regions. For the original DLA model in a square lattices, the fourth harmonic continuously increases as new particles are added to the cluster \cite{Ball}, whereas a null value is found for isotropic patterns. In Fig. \ref{dlascr6}, $\langle \cos 4\theta \rangle$ is plotted as a function of the number of particles. For smaller $\delta$ values ($\delta \le 16 $), the clusters are initially isotropic, i. e., $\langle \cos 4\theta \rangle \approx 0$, but an effective increasing anisotropy emerges during the cluster growth. Such finds agree with the results obtained for the original DLA model \cite{Ball}. Nevertheless, concerning our simulations, for larger $\delta$ values ($\delta \ge 18$) none effective anisotropy were measured. One can argue that this result may be a consequence of our computational limitations, i. .e, it did not allow us to generate sufficiently large clusters for which a measurable anisotropy is present. However, we conjecture that these patterns are isotropic since the characteristic size for which the lattice anisotropy emerges grows continuously up to $\delta \approx 16$ and suddenly diverges for $\delta > 18$. Moreover, the clusters scale as the off-lattice DLA model in the asymptotic limit, as shown in Fig. \ref{dlascr3}, reinforcing the idea that lattice effects are not present. The physical origin of anisotropy break is related to the relaxation process of the particles when they reach the inner regions of the cluster.

\section{Conclusions}
\label{conclu}

In the present work, it was studied an extension of the diffusion-limited-aggregation (DLA) model, in which the screening effects were directly considered. The unitary steps used in  the random walks of the original DLA model were replaced by random flights of fixed length $\delta$. An increasing $\delta$ leads to a decreasing screening for the penetration of the particles in the inner regions of the clusters. 

The model generates DLA-like structures for small $\delta$ values. However, for larger $\delta$ values the patterns have the Eden model scaling properties on smaller length scales but scale as the DLA model on longer length scales. The compact-to-fractal transition was characterized by the radius of gyration $R_g$ and the number of particles on the border of the cluster $S$. Both quantities exhibit two neat scaling regimes separated by a characteristic number of particles $N_{\times}$, i. e., the patterns scale as the Eden model for $N\ll N_{\times}$ but scale as the DLA model for $N\gg N_{\times}$. From a scaling analysis we found that $N_{\times}\sim \delta^{-\alpha}$, $S\sim \delta^{-z_1}$ and $R_g\sim \delta^{-z_2}$  with $\alpha\approx 3.00$, $z_1 \approx 1.50$ and $z_2\approx 0.249$.

When $N\approx N_{\times}$, the patterns exhibit a quasi-periodic finger-like structure characterized by the selection of a principal and other additional lower-amplitude modes in the correspondent Fourier spectra of the cluster boundary. In turn, in the  asymptotic DLA-like scaling regime all wavelengths are present and, consequently, a self-affine boundary, with Hurst exponent $H\approx 0.68$ apparently independent on $\delta$, was observed. Moreover, the simulations strongly suggest that the lattice effects are not present for large $\delta$ values.

\textbf{Acknowledgments}: I thank to M. L. Martins for extensive discussion and valuable comments on the manuscript. This work was partially supported by the CNPq and FA\-PE\-MIG - Brazilian agencies.

\end{document}